\documentclass[prb,twocolumn,showpacs,amsmath,amssymb,superscriptaddress,floatfix]{revtex4}

\usepackage[english]{babel}
\usepackage{amsfonts}
\usepackage{graphicx}
\usepackage{times}

\begin{document}

\title{Optical conductivity  in the $t$-$J$-Holstein Model}

\author{L. Vidmar}
\affiliation{J. Stefan Institute, 1000 Ljubljana, Slovenia}

\author{J. \surname{Bon\v ca}}
\affiliation{Faculty of Mathematics and Physics, University of Ljubljana, 1000
Ljubljana, Slovenia}
\affiliation{J. Stefan Institute, 1000 Ljubljana, Slovenia}

\author{S. \surname{Maekawa}}
\affiliation{Institute for Materials Research, Tohoku University,
Sendai 980-8577, Japan}
\affiliation{CREST, Japan Science and
Technology Agency (JST), Kawaguchi, Saitama 332-0012, Japan}



\date{\today}
\begin{abstract}
Using recently developed numerical method we compute charge stiffness and optical
conductivity of the $t$-$J$ model coupled to optical phonons.  Coherent hole
motion is most strongly influenced by the electron-phonon  coupling within the
physically relevant regime of the exchange interaction. We find unusual
non-monotonous  dependence of the charge stiffness as a function of the exchange
coupling  near the crossover to the strong electron-phonon coupling regime.
Optical conductivity in this regime  shows a two-peak structure.   The
low-frequency peak represents local magnetic excitation, attached to the hole,
while the higher-frequency peak corresponds to the mid infrared band that
originates from coupling to spin-wave excitations, broadened and renormalized by
phonon excitations. We observe no separate peak at or slightly above the phonon
frequency. This finding suggests that  the two peak structure seen  in recent
optical measurements is due to magnetic excitations coupled to lattice degrees of
freedom via doped charge carriers.
\end{abstract}

 \maketitle

\section{Introduction}

Despite many years of intensive research of transport properties of a hole doped
in an antiferromagnetic  background  the proper description of this system
remains a challenging  theoretical problem. The transport
of a doped hole leaves in its wake locally distorted, slowly relaxing  spin
background,   leading to the formation of a dressed quasiparticle  with an
enhanced  effective mass and renormalized  charge stiffness - a measure of a
coherent, free particle like transport. Addition of lattice degrees of freedom to
this already elaborate problem  reflects  the current scientific
interest in the field of correlated electron systems.

Long after the pioneering work \cite{ramsak2}, the  enhanced   interest in
correlated models, coupled to lattice degrees of freedom is primarily fuelled by
experimental  evidence given in part  by  angular resolved photoemission  data
demonstrating that strong  electron-phonon (EP) interaction plays an important
role in low-energy physics of high-$T_c$ materials
\cite{meevasana,rosch,zhou1,alexandrov}. Moreover, recent estimates of transport
based on the pure  the $t-J$ model \cite{gunnarRMD} yield substantially smaller
resistivity in comparison to experiments in the low-doping regime of
La$_{2-x}$Sr$_x$CuO$_4$, that can be explained as a lack of additional lattice
degrees of freedom.

Recent numerical methods  investigating optical conductivity (OC) in correlated
electron systems and systems, where electrons are coupled to bosonic degrees of
freedom, have been focused on the generalized $t-J$ model
\cite{tohyama01,peteropt}, Holstein and generalized electron-boson model
\cite{fehske1,fehske2}, while investigations of the $t-J$-Holstein model have been
until recently  limited to small clusters with 10 sites \cite{fehske3}.
Diagrammatic Quantum Monte Carlo (DMC) method has been applied to resolve OC  of
the  Fr\" ohlich polaron \cite{mishchenkoFROL} and recently also the
$t-J$-Holstein model \cite{mishchenkoOPT}. In the later work authors report on a
two-peak structure in the optical response where the low-$\omega$ peak is due to
polaronic effects while the peak at higher $\omega$  is due to magnetic
excitations, renormalized by lattice degrees of freedom. DMC method does  not
reproduce the well  known string states seen in the spectral function \cite{mukhin} neither the signature  of local magnetic
excitations in optical conductivity  since authors use the self consistent Born approximation (SCBA)
without magnon-magnon vertex corrections for treatig spin degrees of freedom.
Their result seems to contradict calculations based on the dynamical mean-field
theory (DMFT) where a two peak structure is seen in OC only at small $J/t\lesssim
0.3$ \cite{cappelluti}.  In this work authors also show that the low-$\omega$
peak is of the magnetic origin while the higher-$\omega$ peak represents the broad
polaronic band.

\section{Model and Numerical  Method}
The main goal of  this work is to  investigate in depth  optical properties of
the $t$-$J$-Holstein model for the case  of a single hole in the antiferromagnetic
background. We first define the $t$-$J$-Holstein  model on a square lattice
\begin{eqnarray}
H&=& -t\sum_{\langle {i,j}\rangle,s}\tilde c^\dagger_{{i},s} \tilde c_{{j},s} +
J\sum_{\langle i,j\rangle }{\bf S}_{i} {\bf S}_{j},\nonumber \\
 &+&g\sum_{i}(1-n_{i})(a_{i}^+ + a_{i})+
\omega_0\sum_{i}a_{i}^+  a_{i}, \label{ham}
\end{eqnarray}
where $\tilde c_{{i},s}=c_{{i},s}(1-n_{{i},-s})$ is a fermion operator, projected
onto a space of no double occupancy, $t$ represents nearest- neighbor overlap
integral, the sum $\langle i,j\rangle$ runs over pairs of nearest neighbors,
$a_{i}$ are  phonon annihilation operators and $n_{i}=\sum_s n_{{i},s}$. The third
term represents EP coupling $g=\sqrt{8\lambda \omega_0t}$, where
$\lambda$ is the dimensionless EP coupling constant,    and the last term
represents the energy of Einstein phonons $\omega_0$.

We use  recently developed method based on the exact diagonalization within the
limited functional space (EDLFS). \cite{bonca3,bonca2} Since details of the method
have been published elsewhere, \cite{bonca1,bonca2,bonca3} we now briefly discuss
only the main steps of the method. We first construct the limited functional space
by starting from a N\'{e}el state with one hole with a given momentum $\bf k$ and
zero phonon degrees of freedom $\vert \phi_{\bf k}^{(0,0)}\rangle = c_{\bf
k}|{\mathrm{Neel}};0\rangle$, and applying  the generator of states $ \{\vert
\phi_{{\bf k}l}^{(N_h,M)}\rangle \}=(H_{\mathrm{kin}}+H_g^M)^{N_h}\vert \phi_{\bf
k}^{(0,0)}\rangle, $ where $H_{\mathrm{kin}}$  and $H_g$ represent  the first and
the third term respectively of Eq.~\ref{ham}. This procedure generates
exponentially growing basis space of states, consisting of different shapes of
strings in the vicinity of the hole with maximum lengths given by $N_h$ as well as
phonon quanta that are as well located in the vicinity of the hole,  at a maximal
distance $N_h$. Parameter $M$ provides generation of additional phonon quanta
leading to a maximum  number $N_{\mathrm {ph}}^{max}=M N_h$. Full Hamiltonian
given by Eq.~\ref{ham} is diagonalized within this limited functional space taking
into account the translational symmetry while the continued fraction expansion is
used to obtain dynamical properties of the model. The method treats spin, charge
as well as lattice degrees of freedom on equal footing.

  We define OC per doped hole \cite{maldague}
\begin{eqnarray}
 {\boldsymbol\sigma}(\omega)  =  \frac{i}{\omega^+}
 \left(\langle{\boldsymbol{\tau}}\rangle - \boldsymbol{\chi}(\omega)\right)\label{sigma}\\
 \boldsymbol{\chi}(\omega) = i \int_0^{\infty} e^{i\omega^+ t}
 \langle \left[ \mathbf{j}(t),\mathbf{j}(0)\right]\rangle \mathrm{d}t\label{chi}
\end{eqnarray}
where  $ {\boldsymbol{\tau}}  = \sum_{\langle i,j\rangle,s} t_{ij}
(\textbf{R}_{ij}\otimes\textbf{R}_{ij}) \tilde c_{j,s}^{\dagger}\tilde c_{i,s} $
represents the stress tensor, $\textbf{j}  = i \sum_{\langle i,j\rangle,s} t_{ij}
\textbf{R}_{ij} \tilde c_{j,s}^{\dagger}\tilde c_{i,s}$ is the current operator,
$t_{ij}=-t$ for next nearest neighbors only and zero otherwise, and
$\textbf{R}_{ij}=\textbf{R}_{j}- \textbf{R}_{i}$. We also note that in the case of
next-neighbor tight binding models, $\langle \boldsymbol{\tau}\rangle$ is related
to the kinetic energy, $\langle \tau_{\mu,\mu}\rangle = -\langle
H_{\mathrm{kin}}\rangle/2$.

\section{Charge Stiffness and sum-rules}

Charge stiffness per doped hole can be on a square lattice for the
$t$-$J$-Holstein model   computed via its spectral representation \cite{shastry}
\begin{eqnarray}
 D_{\mu,\mu} &=& -{1\over 4} \langle 0\vert
 H_{\mathrm kin}\vert 0 \rangle +  \sum_{n}
\frac{\langle 0|j_{\mu}|n\rangle\langle n|j_{\mu}\vert 0\rangle}{(E_0-E_n)},
\label{dmumu}\\
D_{\mu,\mu} &=&  S^{tot} - S_{\mu,\mu}^{reg},\label{sumrule}
\end{eqnarray}
where $S^{\mathrm{tot}}$ represents normalized optical sum-rule
$\int_{-\infty}^{\infty} \sigma_{\mu,\mu}'(\omega) \mathrm{d}\omega = 2\pi
S^{\mathrm{tot}}$,  while $S^{\textrm reg}$ is defined by $\int_{0^+}^{\infty}
\sigma_{\mu,\mu}'(\omega) \mathrm{d}\omega = \pi S^{\mathrm{reg}}_{\mu,\mu}$ and
${\boldsymbol \sigma'}(\omega)$ represents the real part of the optical
conductivity tensor in Eq.\ref{sigma}.
 We have computed $D_{\mu,\mu}$ in the single-hole ground-state,
{\it i.e.} at $\textbf{k}=(\pm\pi/2,\pm \pi/2)$. It is well know that the
dispersion $E(\textbf{k})$ is highly anisotropic around its single-hole minimum,
which is in turn reflected in the anisotropy of the effective mass tensor
\cite{ramsak2,ramsak1,peteropt}. It is thus instructive to compute tensors
representing the charge stiffness as well as the OC  in the
direction of their eigen-axis, {\it i.e.} along the nodal ($ (\pi/2,\pi/2) \to
(0,0)$) direction  that gives $D_\parallel$,$\sigma_\parallel(\omega)$,  and along
the anti-nodal ($ (\pi/2,\pi/2) \to (\pi,0)$) direction  that leads to $D_\perp$,
and $\sigma_\perp(\omega)$.

\begin{figure}[htb]
\includegraphics[width=7cm,clip]{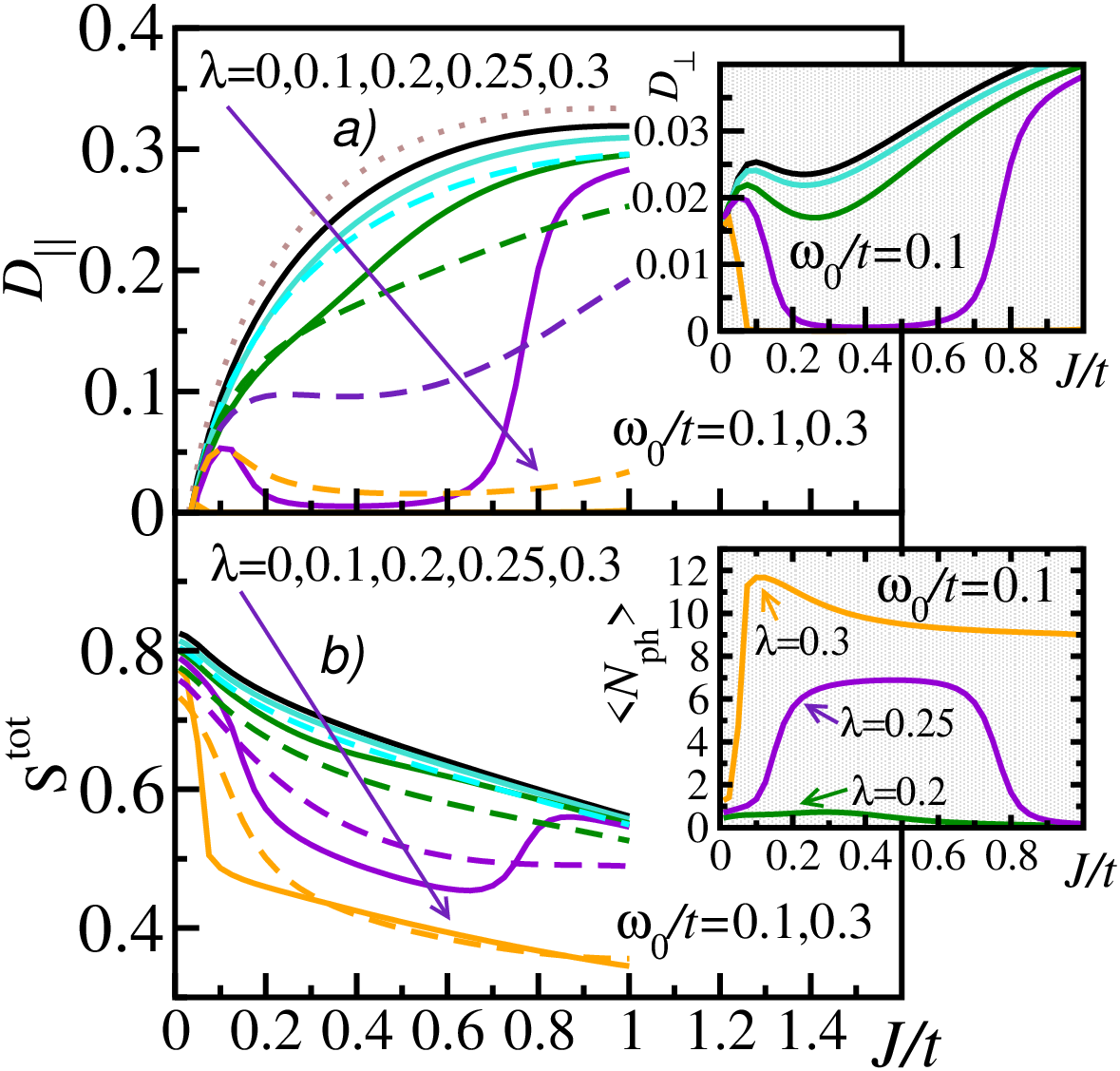}
\caption{(Color online) a) Charge stiffness $D_\parallel$ ( $D_\perp$ in the
inset), b) optical sum-rule $S^{\mathrm{tot}}$
($\langle N_{\mathrm {ph}}\rangle =\sum_i a_i^\dagger a_i$ in the insert)
vs. $J/t$ at  $\omega_0/t=0.1$ (full lines) and $\omega_0/t=0.3$ (dashed lines).
In this and all subsequent figures (except for the dotted line in  a) or else
otherwise indicated) we used: $N_h=8$,  $M=7$, and $N_{\mathrm{st}}\sim 9\times
10^6$.
Dotted  line in a) was for
$\lambda=0$ obtained with states with zero phonon degrees of freedom and the
following set of parameters: $N_h=14$,  $M=0$, and $N_{\mathrm{st}}\sim 5\times
10^6$.
}\label{fig1}
\end{figure}

In Fig.~\ref{fig1}(a) we present the charge stiffness vs. $J/t$ for various values
of EP coupling strength. To obtain accurate results in the strong EP coupling (SC)
limit, we had to rely on only $N_{\mathrm{st}}=9786$ different combinations  of
spin-flip states, while the total number of states, including phonon degrees of
freedom, was $N_{\mathrm{st}}=9\times 10^6$. To test the quality of $\lambda=0$
results, we show with the dashed line $D_\parallel$ computed  with zero phonon
degrees of freedom using $N_{\mathrm{st}}=5\times 10^6$. Agreement with the
$\lambda=0$ case, obtained with $N_{\mathrm{st}}=9786$ is rather surprising, given
the fact that  results were computed using Hilbert spaces that differ nearly three
orders of magnitude. This fast convergence is  in  contrast to calculations on
finite-size clusters where  due to the existence of persistent currents $D$ varies
rather uncontrollably between different system sizes \cite{prelovsek}.

Exploring further  $\lambda=0$ results we observe $D_\parallel\sim 0$  at $J/t\sim
0$ indicating strong scattering on spin degrees of freedom. With increasing $J/t$
$D_\parallel$ steeply increases and around  $J/t\sim 1$ reaches at
$D_\parallel\sim 0.3$ a broad maximum that as well coincides with the maximum of
the bandwidth $W$\cite{bonca2}. In contrast,   the optical sum rule
$S^{\mathrm{tot}}=-\langle H_{\mathrm{kin}}\rangle/4$  monotonically decreases in
the range $0\lesssim J/t \lesssim 1.0$ with increasing $J/t$, Fig.~\ref{fig1}(b).
Due to strong anisotropy in $E({\bf k})$, $D_{\perp}$ remains nearly an order of
magnitude smaller than $D_\parallel$ for $J/t\gtrsim 0.2$ (see the inset of
Fig.~\ref{fig1}(a)).

Turning to finite $\lambda$, $D_\parallel$ expectedly decreases, due to additional
scattering on lattice degrees of freedom. The effect of $\lambda$ on the value of
$D_\parallel$ however varies with $J/t$. This is best seen in the case of $\lambda
= 0.25$ and $\omega_0/t=0.1$ where $D_\parallel$ is  approximately equal to its
$\lambda=0$ value for $J/t\lesssim 0.1$, it then decreases with increasing $J/t$,
reaching its minimum value around $J/t\sim 0.5$ and finally, for larger values of
$J/t\gtrsim0.8$, steeply increases. This non-monotonous begavior is as well
reflected in the bell shaped average phonon number $\langle N_{\mathrm
{ph}}\rangle$ vs. $J/t$, presented in the insert of  Fig.~\ref{fig1}(b). This
behavior is also consistent with the  non-monotonous  functional dependence of
$\lambda_c(J/t)$, representing the  crossover EP coupling strength  to the SC
regime,   Refs.~\cite{peter,bonca3}.

We now make some general comments about the effect of the EP interaction on the
correlated system at the onset of the SC regime. At small values  of $J/t\lesssim
0.1$  EP coupling is less effective, which  seems to be in  contrast to naive
expectations.   We attribute this {\it disentanglement } from lattice degrees of
freedom to the increase of the kinetic energy and the vicinity of the Nagaoka
regime. This effect  particularly evident from the $J/t$-dependence of the average
phonon number $\langle N_{\mathrm {ph}}\rangle$  at $\lambda=0.3$ (see the insert
of Fig.~\ref{fig1}(b)) where an increase followed by a sharp drop of $\langle
N_{\mathrm {ph}}\rangle$ is seen with lowering of $J/t$.  
At the onset of the SC regime, {\it i.e.}  at $\lambda\sim 0.25$, EP coupling is most effective in the physically relevant $J/t\sim 0.3-0.4$ regime, where there is a strong competition between kinetic energy and magnetic excitations.
The critical $\lambda_c$ as
well  reaches its minimum around $J/t\sim 0.3$ as shown in Ref.\cite{bonca3}.  At
larger $J/t\sim 1$ EP coupling becomes  again less effective due to more coherent
quasiparticle motion as reflected in the  enhanced charge stiffess, quasiparticle
weight, as well as the bandwidth\cite{bonca3,bonca2}.

The optical sum-rule $S^{\mathrm{tot}}$, presented in Fig.~\ref{fig1}(b), as well
decreases with increasing $\lambda$. It however remains finite even deep in the SC
regime where $D_\parallel\sim 0$ since
$S^{\mathrm{tot}}$ includes both coherent as well as incoherent transport. The latter
remains finite due to  processes, where the hole hops back and forth
between neighboring sites while leaving lattice deformation unchanged.
Despite charge localization we thus expect nonzero optical response
$\sigma(\omega)$ even deep in  the SC regime, with its spectral weight shifted towards larger
$\omega$ and zero contribution at $\omega=0$. Due to localization we also expect
OC to be isotropic in the SC regime, {\it i.e.}
$\sigma_{\parallel}(\omega)\sim \sigma_{\perp}(\omega)$.

\begin{figure}[htb]
\includegraphics[width=7cm,clip]{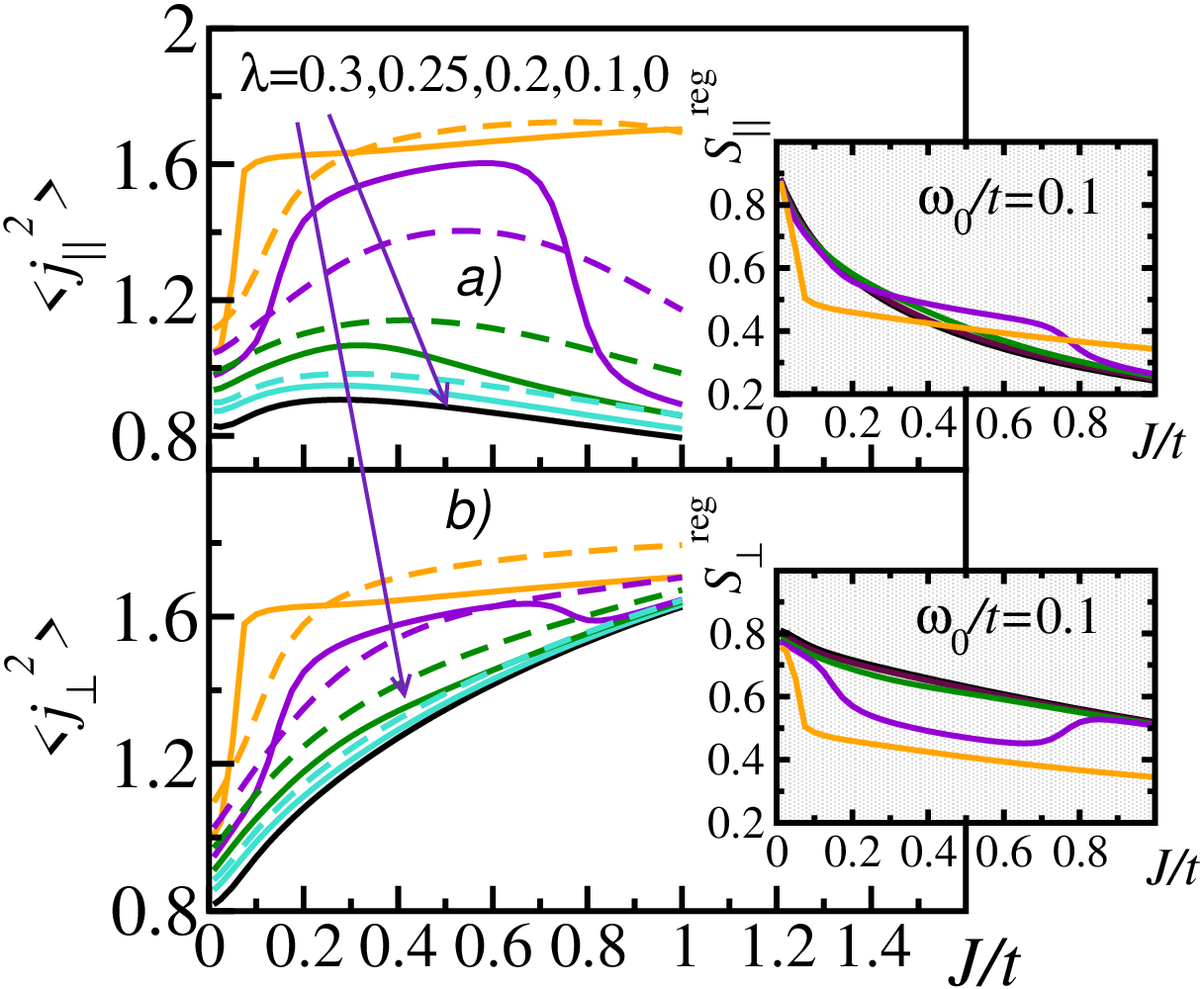}
\caption{(Color online) a) Expectation value of the square of the electrical
current along the nodal direction $\langle j_{\parallel}^2\rangle$
($S^{\mathrm{reg}}_{\parallel}$  in the inset), b) expectation value of the square
of the electrical current along the anti-nodal direction $\langle
j_{\perp}^2\rangle$ ($S^{\mathrm{reg}}_{\perp}$  in the inset) vs. $J/t$ at
$\omega_0/t=0.1$ (full lines) and $\omega_0/t=0.3$ (dashed lines).
}\label{fig2}
\end{figure}

In the insets of Fig.~\ref{fig2} we show $S^{\mathrm{reg}}_{\parallel}$ and
$S^{\mathrm{reg}}_{\perp}$ representing  the integrated regular part of the OC.
$S^{\mathrm{reg}}_{\parallel}$ steeply decreases with increasing $J/t$ due to the
simultaneous increase of coherent transport, captured   by $D_{\parallel}$,  as
well as due to  decrease of $S^{\mathrm{tot}}$, see also Eq.~\ref{sumrule}. We
observe, that EP coupling have little effect on $S^{\mathrm{reg}}_{\parallel}$ for
$J/t\lesssim 0.4$,  since its value is rather independent on $\lambda$ except deep
in the SC regime, {\it i.e.} at $\lambda = 0.3$ in this particular case.  Due to
small values of $D_\perp$ we find $S^{\mathrm{reg}}_{\perp}\sim S^{\mathrm{tot}}$,
see Figs.~\ref{fig1} and ~\ref{fig2} .

In Fig.~\ref{fig2} we present the average of the square  of the electrical current
defining the following sum-rule  $\int_{0^+}^{\infty} \omega
\sigma_{\mu,\mu}'(\omega) \mathrm{d}\omega = \pi \langle j_{\mu,\mu}^2\rangle$,
that furthermore  represents the fluctuation of the current operator. In the
ground state there is no persistent currents that usually appear on finite-size
clusters,   since our method is defined on an infinite lattice. This enables more
reliable calculation of the charge stiffness. At $\lambda=0$ $\langle
j_{\parallel}^2\rangle$ (in Fig.~\ref{fig2}(a))  and $\langle j_{\perp}^2\rangle$
(in Fig.~\ref{fig2}(b)) display rather distinctive $J/t$ dependence. While
$\langle j_{\parallel}^2\rangle$ shows weak non-monotonous dependence on $J/t$,
$\langle j_{\perp}^2\rangle$ shows a substantial increase. With increasing
$\lambda$ current fluctuations as well increase in both directions even though the
increase is more pronounced in the case of $\langle j_{\parallel}^2\rangle$. In
the SC regime we obtain $\langle j_{\parallel}^2\rangle\sim \langle
j_{\perp}^2\rangle$ as a consequence of localization due to lattice degrees of
freedom.

\begin{figure}[htb]
\includegraphics[width=7cm,clip]{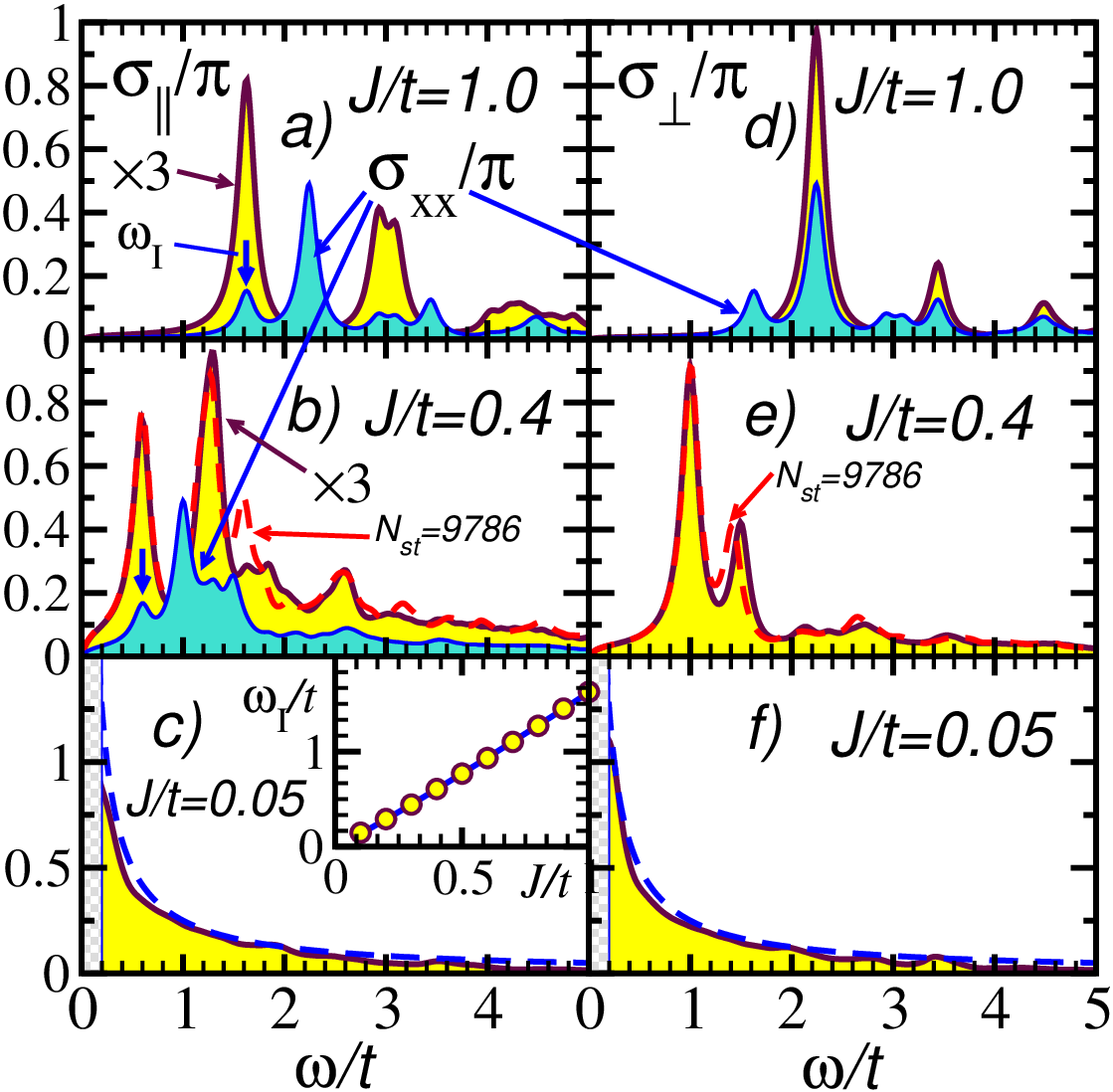}
\caption{(Color online)  $\sigma_{\parallel}$  in a), b), and c), and
$\sigma_{\perp}$ in d), e),  and f)  for three different values of $J/t$ as
indicated in the figures for the $t$-$J$ model for a single doped hole at
$\lambda=0$ and ${\bf k}=(\pi/2,\pi/2)$. Hilbert  space with no phonon degrees of
freedom and $N_{\mathrm{st}}=5\times 10^6$ was used in all cases except in  b) and
e) where  for comparison we in addition  present calculations with
$N_{\mathrm{st}}=9786$. In a), b), and d)  we also show
$\sigma_{xx}=(\sigma_{\parallel}+\sigma_{\perp})/2$ using  turquoise (dark grey)
fill. Arrows in a) and b) indicate positions of lowest-energy peaks, $\omega_\mathrm{I}$.  Dashed lines in c) and f) are given by $\sigma_{\mu,\mu}(\omega) =
\pi/(z\omega)$. Insert in c) represents scaling of $\omega_{\mathrm{I}}$ vs. $J/t$.  In this and in the subsequent figures, the Drude peak is not
shown. Artificial broadening $\epsilon = 0.1t$ was used. Dashed areas in c) and f)
delineate small frequency regimes ($\omega/t\lesssim 0.2$) where at $J/t=0.05$
EDLFS does not lead accurate results due to the vicinity of  the Nagoka regime.
}\label{fig3}
\end{figure}

\section{Dynamic Properties}

Turning to dynamic properties we first  establish numerical efficiency of our
method by presenting optical properties of the   $t$-$J$ model. In Fig.~\ref{fig3}
we display different components of the conductivity  tensor
$\sigma_{\mu,\mu}(\omega)$ in the single-hole minimum ${\bf k}=(\pi/2,\pi/2)$,
computed using EDLFS. At physically relevant value $J/t=0.4$ we reproduce well
known features, characteristic of    $\sigma_{xx}(\omega)$: a) in the regime
$1.6J\lesssim \omega\lesssim 2t$ we find peaks forming a rather broad band,
appearing within the well known mid-infrared (MIR) frequency regime, separated
from the Drude peak (not shown) by a gap  of the order of $J$ and b) there is a
broad featureless tail, extending to large frequencies, $\omega\gtrsim 7t$. MIR
peaks for $J/t\gtrsim 0.2$  scale with the exchange coupling $(J/t)^{\eta}$, where
$\eta\sim 1$. We stress that such scaling is consistent with local magnetic
excitations as well as spin waves. Obtained scaling  is however {\it not} consistent with the
string picture where $\eta=2/3$ (see also Ref.~\cite{string}).  At $J/t=0.4$ as
well as at $J/t= 1$, the lowest peak appears at $J/t\sim 1.6$. Location of the
lowest-frequency peak (indicated by arrows in Figs~\ref{fig3}(a) and (b)) is
surprisingly close to the location of the peak in  OC of the $t$-$J_z$ model in
the limit $J_z/t\to \infty$, given by  $\sigma(\omega)\sim {t^2/J_z}\delta(\omega - {3\over
2}J_z)$\cite{poilblanc1}. In this trivial case the peak  appears at the frequency
that  corresponds to the energy (measured from the N\' eel state) of a single
spin-flip, attached to the hole, created as the hole hops one lattice site from
its origin in the undisturbed N\' eel background. It is somewhat surprising that such a naive
interpretation seems to survive  even in the (spin) isotropic $t$-$J$ model and at
rather small value of $J/t=0.4$. The scaling of the position of the low-frequency
peak closely follows the following expression $\omega_I = 1.62 (J/t)^{1.08}$, indicated by a dashed line, connecting the circles  shown the insert of Fig.~\ref{fig3}(c). Our
results of OC qualitatively agree with those, obtained on small lattice
systems\cite{poilblanc1}.

When conductivity tensor $\boldsymbol{\sigma}(\omega)$ is computed in its eigen
directions, distinct (incoherent) finite-$\omega$ peaks   are
obtained in the case of $\sigma_{\parallel}(\omega)$ and $\sigma_{\perp}(\omega)$,
as best seen at $J/t=1$ and $J/t=0.4$. For comparison we present in
Figa.~\ref{fig3} (a),(b), and (d)  $\sigma_{xx}(\omega)$, that consists of all the
peaks characteristic for both  $\sigma_{\parallel}(\omega)$ as well as
$\sigma_{\perp}(\omega)$. The reason is, that the ground state at ${\bf
k}=(\pi/2,\pi/2)$ or $\Sigma$- point  belongs to an irreducible representation
$\Sigma_1$ of the small group of ${\bf k}$, {\it i.e.} $C_2$. Current operators
$j_{\parallel}$ and $j_{\perp}$, defining $\sigma_{\parallel}(\omega)$ and
$\sigma_{\perp}(\omega)$ through Eqs.\ref{sigma}, and \ref{chi}  transform as
distinct  irreducible representations $\Sigma_1$ and $\Sigma_2$. Selection rules
allow only transitions into states that transform according to a direct product of
irreducible representations of the group $C_2$.
Since $j_x$ does not transform
according to irreducible representations of $C_2$, the above mentioned selection
rules do not apply.
\begin{figure}[htb]
\includegraphics[width=7cm,clip]{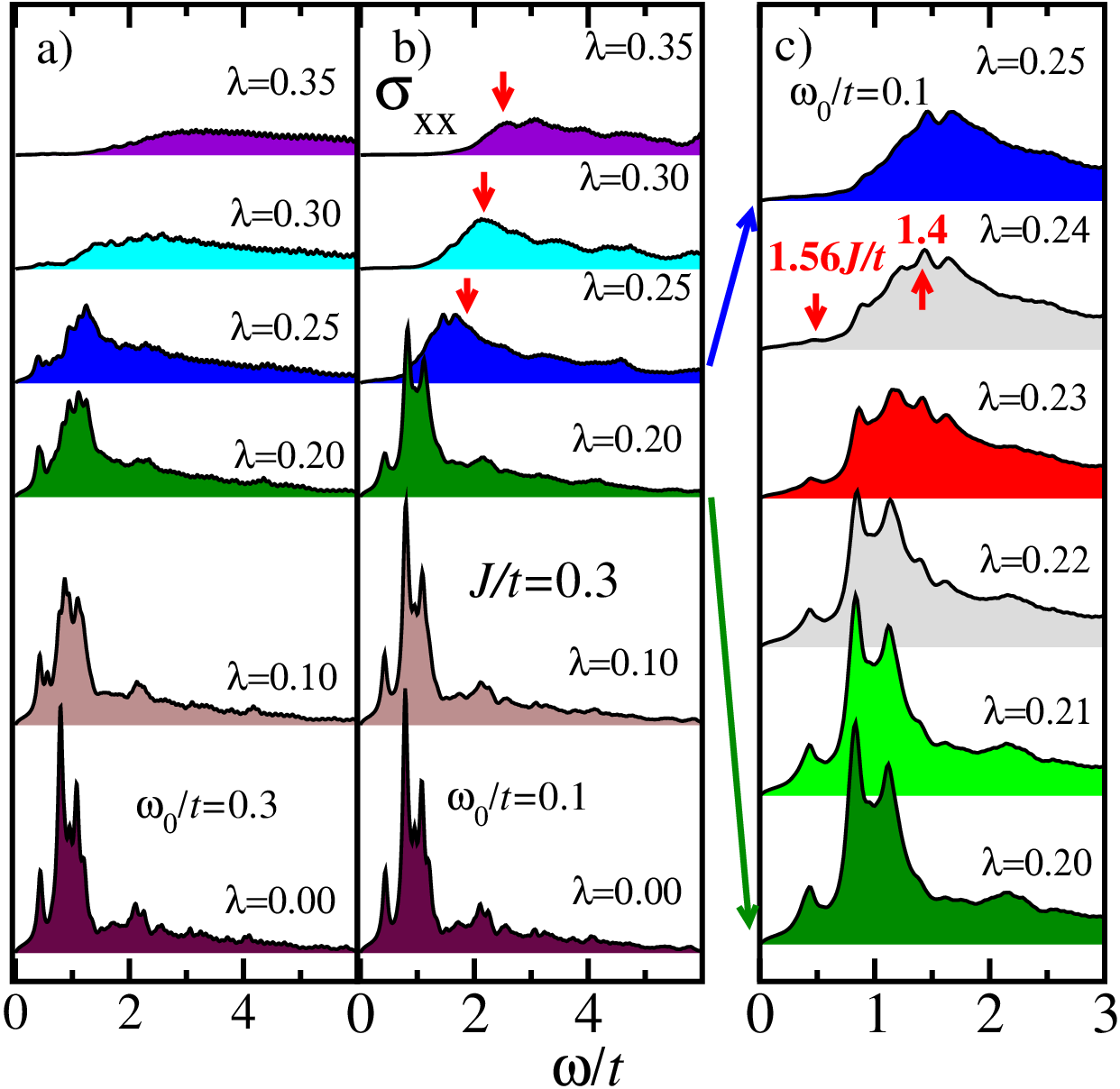}
\caption{(Color online) $\sigma_{xx}$  for $\omega_0/t=0.3$ in a),
$\omega_0/t=0.1$ in b and c) at $J/t=0.3$,  and  at ${\bf k}=(\pi/2,\pi/2)$. Total
number of functions was $N_{\mathrm{st}}=9 \times 10^6$. Up to 56 phonon quanta
was used to obtain accurate results for $\lambda \gtrsim 0.2$.  Arrows in b)
indicate $\omega_{\mathrm{II}}=16\lambda \tilde t$ where $\tilde t = 0.45 t$.
Units of $\sigma$ are arbitrary, yet chosen identical in a) and b); a different
scale was used for c), nevertheless identical among different plots in c).
Artificial broadening was set to $\epsilon/t = 0.05$.
}\label{fig4}
\end{figure}

In Figs.~\ref{fig3} (b) and (e) we present as well results, computed on a much
smaller set of states, {\it i.e.} with $N_{\mathrm{st}} = 9786$. Apart for a small
shift of one of the MIR peaks at larger $\omega$, the agreement with results,
obtained with  more than three orders of magnitude larger systems
($N_{\mathrm{st}} = 5\times 10^6$) underlines the efficiency of our method.
Obtaining relevant results for the pure $t$-$J$ model at moderate number of states
is of crucial importance for successful implementation of additional lattice
degrees of freedom. Last, we present in  Figs.~\ref{fig3}(c) and (f) results at
small $J/t=0.05$. Dashed lines represent known analytical estimate $\sigma(\omega)
= \pi/(z\omega)$, where $z=4$, Ref.\cite{rice}. This result is characteristic for
systems with a nearly constant density of states and diffusive hole motion where
current matrix elements  $\vert\langle 0\vert j_\mu\vert n \rangle\vert $ are
roughly independent of $n$\cite{rice}.  Good agreement with the analytical results
in the small $J/t$ limit  is of particular importance since our method is by
construction,   based on the existence of the long-range  N\' eel order, targeted
to be valid predominantly in the regime of  intermediate to large values of the
exchange constant $J/t$. We also note  that in the limit of small-$J/t$ optical
properties for $\omega/t\gtrsim J/t$ become isotropic.

\begin{figure}[htb]
\includegraphics[width=7cm,clip]{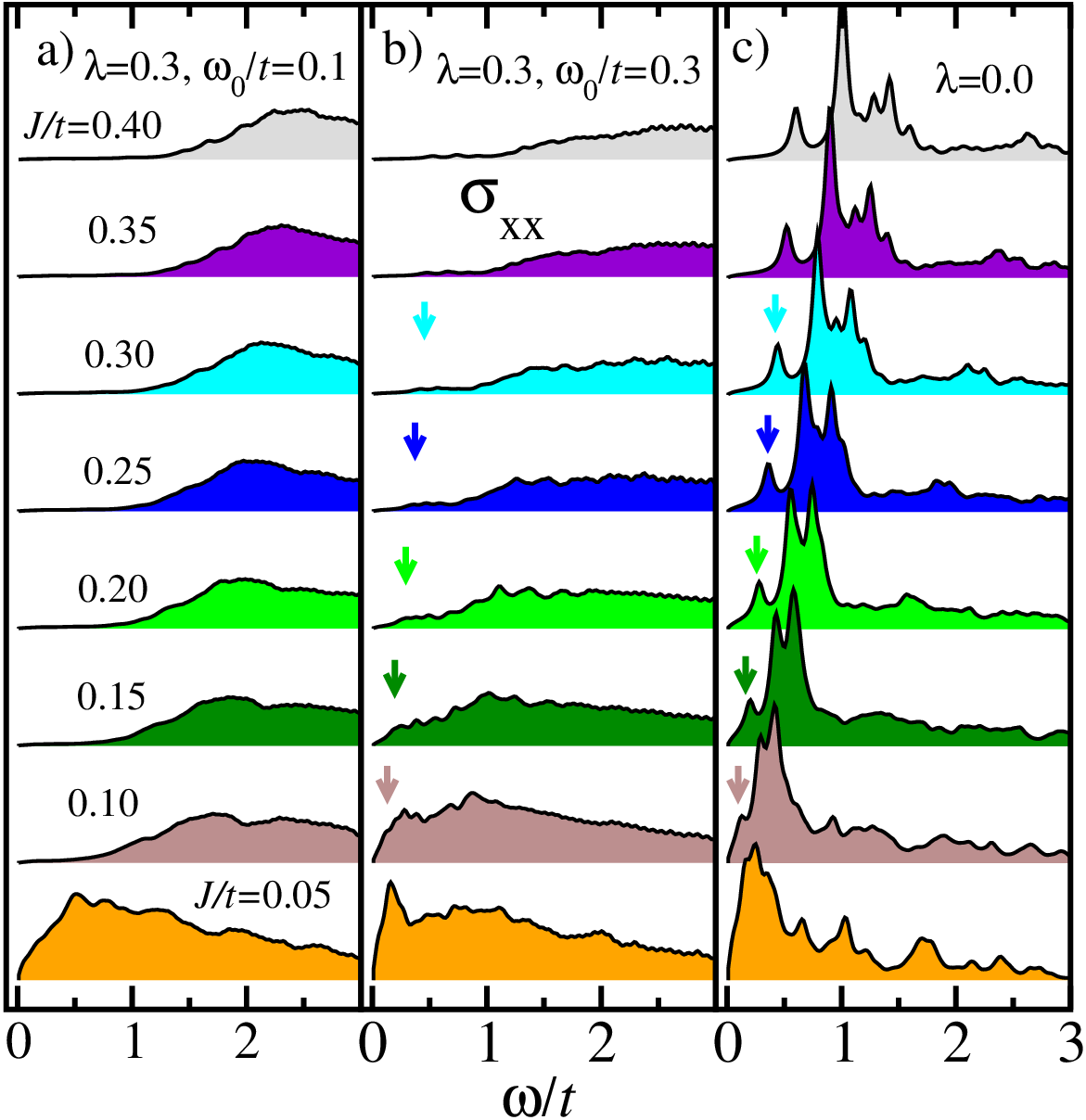}
\caption{(Color online) $\sigma_{xx}$  at $\lambda=0.3$;  a) $\omega_0/t=0.1$, b)
$\omega_0/t=0.3$, and $\lambda=0.0$ in c) at ${\bf k}=(\pi/2,\pi/2)$.  Arrows in
b) and c)  indicate positions of the lowest-frequency state as it appears at
respective values of $J/t$ at $\lambda=0$. Artificial broadening was set to
$\epsilon/t = 0.05$.
}\label{fig5}
\end{figure}

We now focus   on the influence of increasing EP coupling $\lambda$ on optical
properties   of the $t$-$J$ Holstein model in the adiabatic regime, {\it i.e.} for
$\omega_0/t=0.3$,  Fig.~\ref{fig4}(a) and $\omega_0/t=0.1$,   Fig.~\ref{fig4}(b) and (c) the latter
value  being  relevant for cuprates. Increasing EP coupling $\lambda$ leads to
three main effects: a) the spectra  progressively shift towards higher frequencies
while the total spectral weight decreases (see also the inset of
Fig.~\ref{fig3}(b)),
b) magnetic excitations that form a band in the
MIR regime broaden and diminish with increasing $\lambda$;
they finally disappear in the SC
regime where they are replaced by a broad  polaron-like band that clearly
originated from the renormalized MIR peaks. The peak  of the well formed broad
band at $\omega_0/t=0.1$ in the regime $0.25 \lesssim\lambda\lesssim0.4$ roughly
scales with $\omega_{\mathrm{II}}\sim 16\lambda \tilde t$ where $\tilde t=0.45t$
represents renormalized hopping due to EP interaction. At $\omega_0/t=0.3$ a
broader, featureless band is formed, and
c) a large gap opens in the SC regime.

In contrast to numerical results of Ref.~\cite{mishchenkoOPT}, we observe no
separate peak at or slightly above the phonon frequency. This is more clearly seen
in  Fig.~\ref{fig4}(c) where $\sigma_{xx}(\omega)$ is shown in an  expanded
frequency range. This result is consistent with  DMFT calculations of
Ref.~\cite{cappelluti}. Nevertheless,  we find quantitative agreement at
$\lambda\sim 0.24$  with measurements on (Eu$_{1-x}$Ca$_x$)Ba$_2$Cu$_3$O$_6$ in the
low hole-doping regime  published in Ref.~\cite{mishchenkoOPT}. In our calculation $\lambda\sim 0.24$ represents the maximum EP coupling constant where 
the low-$\omega$ peak, located at $\omega_{\mathrm I}\sim 1.56J\sim 187 meV$ (
choosing $t=400  meV$ and $J/t=0.3$), is just barely visible. This peak is, as discussed above,   due to the local magnetic excitation and 
remains separated from the continuum forming the rest of the MIR band.
Experimental value of the corresponding peak is
$\omega_{\mathrm{I}}^{\mathrm{exp}}= 174 meV$~\cite{mishchenkoOPT}. The higher $\omega-$ peak at
$\omega_{\mathrm{II}}\sim 1.4t=560meV$ (experimental value is
$\omega_{\mathrm{II}}^{\mathrm{exp}}= 590 meV$) corresponds to MIR band, slightly
broadened and renormalized by phonon excitations. This part of OC is in agreement with
calculations in Ref.~\cite{mishchenkoOPT}. 


Our explanation of the experimental results relies on the conjecture that lightly
doped  (Eu$_{1-x}$Ca$_x$)Ba$_2$Cu$_3$O$_6$ compound lies in the crossover from
from weak to strong coupling electron-phonon regime where physical properties
(quasiparticle weight, charge stiffness and dynamic properties) are extremely
sensitive to small changes of $\lambda$. This is evident from Fig.~\ref{fig4} and
from results, published in  Ref.~\cite{bonca3}. MIR peak in OC is at $\lambda=0$
centered around $\omega_{\mathrm{II}}=2J=240 meV$. This value corresponds to the
peak of the magnon density of states\cite{mukhin} it however underestimates the position of the
main peak, seen in the experiment of Ref.~\cite{mishchenkoOPT}. Increasing
$\lambda$ beyond the weak coupling regime $\lambda>\lambda_c$, the center of MIR
peaks starts moving towards higher frequencies and broadens as it transforms into
a wide polaron band, thus approaching the experimental value. Simultaneously the
peak due to the local magnetic excitation at $\omega_{\mathrm{I}}$ as well
broadens and disappears above $\lambda\gtrsim 0.24$.


The lack of a  peak at $\omega\sim\omega_0$ in OC can be explained in simple terms
in the large-$J_z/t$ limit of the simplified $t$-$J_z$-Holstein model.  Starting
from a hole in the N\' eel background, the lowest energy contribution to
$\sigma_{xx}(\omega)$ comes from the hop of the hole to the neighboring site. This
move generates a single spin-flip with the energy $E_1=3J_z/2$ above the ground
state.  The contribution to OC that would include a single phonon excitation would
thus be located at $\omega\gtrsim 3J_z/2+\omega_0$.

In order to explore the  interplay of magnetic and polaronic degrees of freedom in the structure of $\sigma_{xx}(\omega)$ in   more detail, we present in Figs.~\ref{fig5}(a) and (b) comparison of optical spectra at fixed $\lambda=0.3$ and  different values of the exchange interaction $J/t$. Decreasing  $J/t$ leads
to a shift of the broad  polaronic peak towards smaller values of $\omega$. At
smaller $\omega_0/t=0.1$ more pronounced structure abruptly appears  at low
$\omega/t\lesssim 0.5$  at small $J/t=0.05$, Fig.~\ref{fig5}(a). At larger value
of $\omega_0/t=0.3$, Fig.~\ref{fig5}(b), a shoulder starts appearing at $J/t=0.3$
in the low-$\omega$ regime that corresponds  to  the onset  of the respective
magnetic peaks (as indicated by arrows in Fig.~\ref{fig5}(c)) of the pure $t$-$J$
model.  Below $J/t\lesssim 0.2$ well formed peaks emerge  being clearly of the
magnetic origin. The disentanglement of lattice degrees of freedom, clearly seen
in Fig.~\ref{fig5}(b), is consistent with  DMFT calcutions\cite{cappelluti}.

\section{Summary}

In summary, we have explored effects of magnetic as well as lattice   degrees o
freedom on optical properties of the $t$-$J$-Holstein model. EDLFS captures well
optical properties of a single hole in the $t$-$J$-Holstein  model in the range of
physically relevant parameters of the model since it treats spin and lattice
degrees of freedom on equal footing. Competition between kinetic energy and spin
degrees of freedom strongly influences the coherent hole motion as measured by
charge stiffness  near the crossover to SC polaron regime.  In the adiabatic
regime increasing EP coupling leads to the shift of the OC spectra towards higher
frequencies and  broadening of peaks  that in the pure $t$-$J$ model originate in
magnetic excitations. As an important as well as unusual finding we report a lack
of a peak in the OC spectra at or slightly above the phonon frequency that we
attribute to the inherently strong correlations that are  present in the $t$-$J$
model. This finding suggests that the two peak structure seen  in recent optical
measurements is entirely due to magnetic excitations. Based on our calculations,
the two peak structure can be explained with the observation of local magnetic
excitations, created by the hole motion at lower frequencies and the contribution
of spin waves, coupled via doped hole to lattice degrees of freedom at higher
frequencies.


\acknowledgments J.B. acknowledges stimulating discussions with I. Sega,  A. S.
Mishchenko and the financial support of the SRA under grant P1-0044. J.B.
furthermore acknowledges R. Krivec for his valuable numerical support and
maintaining extremely stable operating system on Sun clusters where all numerical
calculations have been performed. S.M. acknowledge the financial support of the
Next Generation Super Computing Project of Nanoscience Program, CREST, and
Grant-in-Aid for Scientific Research from MEXT.

\bibliography{manuopt}

\begin{thebibliography}{26}
\bibitem[{str()}]{string}
\bibinfo{journal}{String states represent excited states of a particle (hole)
  in the linear potential created by the overturned spins left in the wake of a
  mobile hole as it propagates through an ordered N\' eel background. Since
  spin fluctuation arising from the off-diagonal part of the exchange
  interaction keep erasing the trace of overturned spins, string states, in
  reality, represent only a naive and approximate physical picture of excited
  states of the hole doped in the antiferromagnetic background. The physical
  realization of the string picture emerges as scaling of excited state
  energies with $J/t$ as $\Delta E \sim (J/t)^{\eta}$ where $\eta\sim 2/3$.}
\end{thebibliography}


\end{document}